\documentclass[%
 aip,
 amsmath,amssymb,
 reprint,%
]{revtex4-1}

\usepackage{graphicx}
\usepackage{dcolumn}
\usepackage{bm}
\usepackage[dvipsnames]{xcolor}
\usepackage[normalem]{ulem}
\usepackage[utf8]{inputenc}
\usepackage[T1]{fontenc}
\usepackage{mathptmx}
\usepackage{hyperref}
\usepackage{etoolbox}
\usepackage{soul}

\DeclareMathOperator\erf{erf}

\newcommand{\angstrom}{\mbox{\normalfont\AA}}

\DeclareUnicodeCharacter{2009}{}
\DeclareUnicodeCharacter{2212}{}
\DeclareUnicodeCharacter{0080}{}
\DeclareUnicodeCharacter{0081}{}

\makeatletter
\def\@email#1#2{%
 \endgroup
 \patchcmd{\titleblock@produce}
  {\frontmatter@RRAPformat}
  {\frontmatter@RRAPformat{\produce@RRAP{*#1\href{mailto:#2}{#2}}}\frontmatter@RRAPformat}
  {}{}
}%
\makeatother
\begin{document}
\preprint{AIP/123-QED}

\title{A versatile laser-based apparatus for time-resolved ARPES with micro-scale spatial resolution}

\author{S. K. Y. Dufresne}
 \email{sdufresne@phas.ubc.ca}
\affiliation{Quantum Matter Institute, University of British Columbia, Vancouver, BC V6T 1Z4, Canada} 
\affiliation{Department of Physics \& Astronomy, University of British Columbia, Vancouver, BC V6T 1Z1, Canada} 

\author{S. Zhdanovich}
\affiliation{Quantum Matter Institute, University of British Columbia, Vancouver, BC V6T 1Z4, Canada} 
\affiliation{Department of Physics \& Astronomy, University of British Columbia, Vancouver, BC V6T 1Z1, Canada} 

\author{M. Michiardi}
\affiliation{Quantum Matter Institute, University of British Columbia, Vancouver, BC V6T 1Z4, Canada} 
\affiliation{Department of Physics \& Astronomy, University of British Columbia, Vancouver, BC V6T 1Z1, Canada} 

\author{B. G. Guislain}
\affiliation{Quantum Matter Institute, University of British Columbia, Vancouver, BC V6T 1Z4, Canada} 
\affiliation{Department of Physics \& Astronomy, University of British Columbia, Vancouver, BC V6T 1Z1, Canada}
\author{M. Zonno}
\affiliation{Quantum Matter Institute, University of British Columbia, Vancouver, BC V6T 1Z4, Canada} 
\affiliation{Department of Physics \& Astronomy, University of British Columbia, Vancouver, BC V6T 1Z1, Canada} 
\author{S. Kung}
\affiliation{Quantum Matter Institute, University of British Columbia, Vancouver, BC V6T 1Z4, Canada} 
\affiliation{Department of Physics \& Astronomy, University of British Columbia, Vancouver, BC V6T 1Z1, Canada} 

\author{G. Levy}
\affiliation{Quantum Matter Institute, University of British Columbia, Vancouver, BC V6T 1Z4, Canada} 
\affiliation{Department of Physics \& Astronomy, University of British Columbia, Vancouver, BC V6T 1Z1, Canada} 

\author{A. K. Mills}
\affiliation{Quantum Matter Institute, University of British Columbia, Vancouver, BC V6T 1Z4, Canada} 
\affiliation{Department of Physics \& Astronomy, University of British Columbia, Vancouver, BC V6T 1Z1, Canada} 

\author{F. Boschini}
\affiliation{Quantum Matter Institute, University of British Columbia, Vancouver, BC V6T 1Z4, Canada} 
\affiliation{Centre Energie Materiaux Telecommunications, Institut National de la Recherche Scientifique, Varennes, Quebec J3X 1S2, Canada}

\author{D. J. Jones}
\affiliation{Quantum Matter Institute, University of British Columbia, Vancouver, BC V6T 1Z4, Canada} 
\affiliation{Department of Physics \& Astronomy, University of British Columbia, Vancouver, BC V6T 1Z1, Canada} 

\author{A. Damascelli}
 \email{damascelli@phas.ubc.ca}
\affiliation{Quantum Matter Institute, University of British Columbia, Vancouver, BC V6T 1Z4, Canada} 
\affiliation{Department of Physics \& Astronomy, University of British Columbia, Vancouver, BC V6T 1Z1, Canada} 


\begin{abstract}

We present the development of a versatile apparatus for a 6.2 eV laser-based time and angle-resolved photoemission spectroscopy with micrometer spatial resolution (time-resolved $\mu$-ARPES). With a combination of tunable spatial resolution down to $\sim${11} $\mu$m, high energy resolution ($\sim$11 meV), near-transform-limited temporal resolution ($\sim$280 fs), and tunable 1.55 eV pump fluence up to $\sim$3 mJ/cm$^2$, this time-resolved $\mu$-ARPES system enables the measurement of ultrafast electron dynamics in exfoliated and inhomogeneous materials. We demonstrate the performance of our system by correlating the spectral broadening of the topological surface state of Bi$_2$Se$_3$ with the spatial dimension of the probe pulse, as well as resolving the spatial inhomogeneity contribution to the observed spectral broadening. Finally, after \textit{in-situ} exfoliation, we performed time-resolved $\mu$-ARPES on a $\sim$30 $\mu$m few-layer{-thick} flake of transition metal dichalcogenide WTe$_2$, thus demonstrating the ability to access ultrafast electron dynamics with momentum resolution on micro-exfoliated and twisted materials.

\end{abstract}
\maketitle
\date{\today}
\section{Introduction}

Angle-resolved photoemission spectroscopy (ARPES) is a well-established experimental technique used to directly probe the momentum-resolved electronic structure of crystalline materials. The underlying principle of ARPES is the photoelectric effect, which describes the process of photoemitting electrons via the absorption of high-energy photons. The emitted photoelectrons are then characterized in terms of their kinetic energy and emission angle, providing valuable information about the electronic band structure in the form of spectral intensity $\text{I}(\text{E}_{\text{kin}},\textbf{k})$. Beyond imaging electronic band structure, the spectral intensity encodes information about underlying interactions in strongly correlated systems~\cite{damascelli_probing_2004,sobota_angle-resolved_2021}. For example, in materials with strong electron-phonon coupling, ARPES spectra can exhibit a kink in the dispersion at an energy associated with the phonon mode, revealing information about which modes affect the observed electronic properties~\cite{wen_unveiling_2018,devereaux_anisotropic_2004,cuk_review_2005, ludbrook_evidence_2015}. Likewise, in materials with strong spin-orbit coupling, ARPES spectra can exhibit a characteristic spin-splitting of the bands, which reflects the coupling between the electron spin and orbital momentum in the crystal lattice~\cite{xia_observation_2009}.

While equilibrium ARPES has been instrumental in unveiling one-particle dynamics and interactions in correlated systems, the extension of ARPES into the time domain enables the study of light-induced multi-particle dynamics and out-of-equilibrium phenomena in bulk materials. Time- and angle-resolved photoemission spectroscopy (TR-ARPES) combines ARPES with a pump-probe approach allowing for the simultaneous exploration of temporal, energy, and momentum degrees of freedom (DOF)~\cite{boschini_accepted_2023,smallwood_ultrafast_2016}. This technique has facilitated the observation of unoccupied states~\cite{sobota_ultrafast_2012}, the generation and subsequent decay of phonons~\cite{perfetti_ultrafast_2007,sobota_distinguishing_2014,na_direct_2019}, Bloch Floquet states~\cite{wang_observation_2013,mahmood_selective_2016}, and photoinduced phase transitions~\cite{perfetti_time_2006,okazaki_photo-induced_2018}--observations that were otherwise inaccessible using equilibrium ARPES. 

{More recently, stimulated by advancements in exfoliation techniques and the nascent field of twistronics, which have led to the emergence of interesting physical phenomena and novel phases of matter in the two-dimensional counterparts of bulk materials~\cite{novoselov_electric_2004,novoselov_two-dimensional_2005}, a major effort has been invested in the development of equilibrium ARPES approaches with high spatial resolution, i.e. $\mu$-ARPES and nano-ARPES. These experiments are enabled by light sources capable of tightly focusing the probe photon beam to sub-10 $\mu$m ($\mu$-ARPES) or sub-1 $\mu$m (nano-ARPES), which are capable of probing exfoliated flakes with spatial dimensions on the order of 10 $\mu$m~\cite{cattelan_perspective_2018,mo_angle-resolved_2017}. These techniques have provided a route to answer questions pertaining to the role that dimensionality plays in the momentum-resolved electronic properties of exfoliated materials, heterostructures, and highly inhomogeneous samples. To this end, it is equally desirable to extend also TR-ARPES approaches towards micrometer scale spatial resolution.} 

 \begin{figure*}[t!]
\centering
\includegraphics[width=\textwidth]{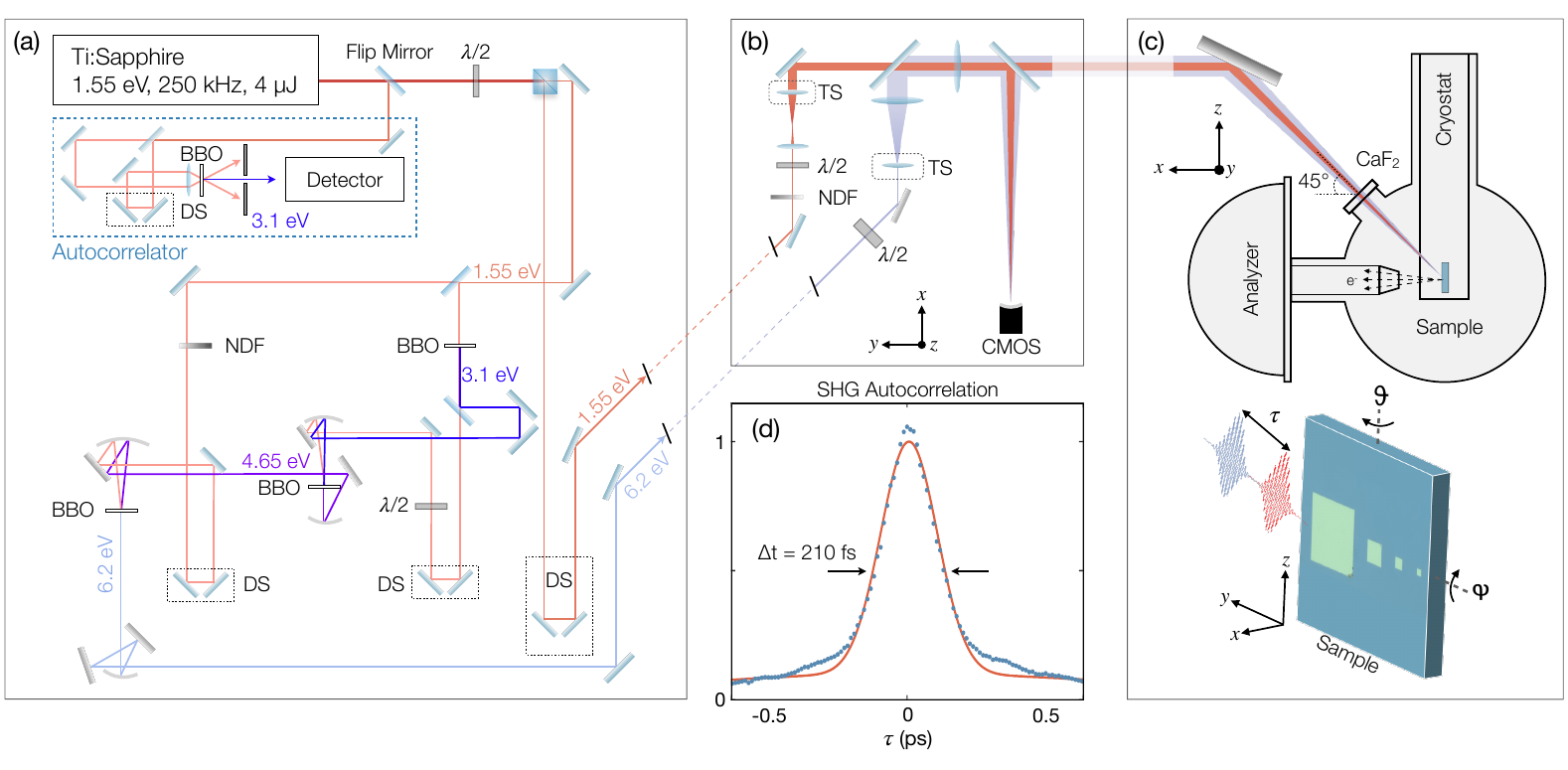}
 \caption{Time-resolved $\mu$-ARPES setup. (a) Schematic layout of the ultrafast Ti:Sapphire source, 6.2 eV generation process, and 1.55 eV pump propagation for TR-ARPES. BBO: $\beta-$BaB$_2$O$_4$; FF: fundamental frequency, corresponding to the source output centered at 800 nm; SHG: second harmonic generation using a 0.5 mm thick BBO crystal cut at $\theta$ = 29.2$^\circ$ for type I phase matching; THG: third harmonic generation 0.2 mm thick BBO cut at $\theta$ = 44.3$^\circ$ {for type I phase matching}; NDF: neutral density filter; FHG: fourth harmonic generation using a 0.1 mm BBO crystal cut at $\theta$ = 64.8$^\circ$ {for type I phase matching}. (b) Schematic layout of the pump and probe expanding and focusing optics. The pump {beam} is {propagated through a NDF and half-waveplate for power and polarization control, respectively, before entering a Keplerian telescope configuration with the second lens mounted on a linear translation stage (TS), thereby controlling the divergence of the pump beam and the spot size incident on the sample.} The probe propagates through a half-waveplate before entering a Galilean expanding telescope, where the first lens is mounted on a {TS} to control the divergence of the probe{, and consequently the spot size incident on the sample}. A dichroic mirror recombines the pump and probe before they propagate co-linearly to the focusing lens and are reflected into the ARPES chamber through a CaF$_2$ optical window at 45$^\circ$ above the sample normal. Approximately 5\% of the pump and probe power is transmitted through a beamsplitter mounted before the CaF$_2$ window, and is incident on a high-resolution CMOS sensor for simultaneous imaging of the pump and probe beam profiles. (c) Schematic showing a transversal cut through the ARPES chamber, highlighting the 45$^\circ$ angle of incidence of the probe with respect to the sample {(Si/SiO$_2$ wafer patterned with Au squares of decreasing side, from 300 - 30 $\mu$m)}. The sample orientation is indicated by the axis labels where $x$ controls the distance between the sample and the entrance slit, while $y$ and $z$ control the horizontal and vertical position of the sample at a fixed distance from the analyzer, respectively. (d) {Cross-correlation of two 1.55 eV pulses using an SHG autocorrelator, showing a width of $\sim$210 fs determined by a Gaussian fit.}}
 \label{Fig:canopy}
 \end{figure*}

Time-resolved $\mu$-ARPES enables the tracking of light-induced electron dynamics in exfoliated and inhomogeneous materials: while an intense pump pulse optically excites the system from its ground state, a subsequent micro-meter sized probe pulse photoemits the excited electrons. A lower photon energy source in the ultraviolet (UV) regime between 5.9 - 7 eV is typically generated from nonlinear optical crystals such as $\beta-$BaB$_2$O$_4$ (BBO) or KBe$_2$BO$_3$F$_2$ (KBBF) ~\cite{koralek_experimental_2007,faure_full_2012,ishida_time-resolved_2014,yang_time-_2019,bao_ultrafast_2022}, while higher photon energy sources in the extreme UV regime, capable of reaching photon energies exceeding 20 eV, are typically generated via high-harmonic generation (HHG) in a gas~\cite{mills_cavity-enhanced_2019,sie_time-resolved_2019,puppin_time-_2019,keunecke_time-resolved_2020,lee_high_2020}. Laser-based systems capable of achieving micron-scale spatial resolution typically use a combination of external expanding telescope schemes and focusing optics mounted in vacuum to achieve spatial resolutions better than 5 $\mu$m ~\cite{iwasawa_development_2017,xu_development_2023}. Moreover, a time-resolved $\mu$-ARPES system should encompass a mechanically-stable sample manipulator with cryogenic cooling capabilities, diagnostic tools to characterize the spot size and pointing stability of the laser pulses, as well as a high-resolution camera to locate and image exfoliated flakes in front of the electron detector.

In this work, we describe the development of a 1.55 eV pump and 6.2 eV probe source for time-resolved $\mu$-ARPES with tunable spatial resolution down to $\sim$11 $\mu$m (Section~\ref{ExperimentalSetup}) and demonstrate the performance of our system by{:} (1) {resolving the surface inhomogeneities that are responsible for increased spectral broadening in the topological insulator Bi$_2$Se$_3$} (Section ~\ref{Bi2Se3}){; and} (2) measuring the {time-resolved} electronic properties of an \textit{in-situ} exfoliated sub-30 $\mu$m WTe$_2$ flake on Ag(111) (Section~\ref{WTe2}).

\section{Experimental Setup}\label{ExperimentalSetup}
\subsection{System Information}\label{SystemInfo}
ARPES measurements are performed in the UBC-Moore Center for Ultrafast Quantum Matter equipped with a SCIENTA-Omicron DA30-L high-resolution hemispherical electron analyzer, in a UHV chamber with base pressures lower than 5$\times$ 10$^{-11}$ Torr. The spectrometer is equipped with a helium-flow cryogenic manipulator, providing stable temperature control of the samples to sub-10 K temperatures. The manipulator possesses five motorized DOF, which consist of two rotational ($\theta, \phi$) DOF to orientate the sample, and three translational DOF (\textit{x,y,z}) [see Fig.~\ref{Fig:canopy}(c)].
 
The vibration of the cryostat and, as a result, of the sample position follows a Gaussian distribution {$F(x,z) \propto \exp{(-\frac{y^2}{2\delta_{y}^2}-\frac{z^2}{2\delta_{z}^2})}$, where $\delta_{y}$ and $\delta_{z}$ are the standard deviations in the sample position measured to be $\sim$4.6 $\mu$m and $\sim$3.3 $\mu$m in $y$ and $z$, respectively}. On the spatial scale of the tightly-focused probe, these variations cannot be neglected and will contribute to the total spatial resolution of our apparatus, as will be discussed in Section~\ref{SpatialResolution}. 
 
The system also features a load-lock {and preparation chamber} with the second stage equipped with {a sputter gun} and e-beam heater for sputtering-annealing of samples.

\subsection{Light Source} \label{LightSource}
A schematic of the 6.2 eV generation process and 1.55 eV pump propagation is shown in Fig.~\ref{Fig:canopy}(a). The ultrafast light source is based on a Ti:sapphire laser (Vitesse Duo and regenerative amplifier RegA 9000 by Coherent), generating $\sim$150 fs pulses [characterized from the cross-correlation of two 1.55 eV pulses using a second harmonic generation (SHG) autocorrelator, as shown in Fig.~\ref{Fig:canopy}(d)] at a central wavelength of 800 nm  at a repetition rate of 250 kHz, and an output pulse energy of $\sim$4 $\mu$J. The source output is separated into two branches which form the basis of both the 6.2 eV probe and the 1.55 eV pump. A portion of 1.55 eV output is focused onto a BBO crystal for {SHG} producing {3.1 eV}, which is recombined with the residual 1.55 eV on a second BBO for third harmonic generation (THG), producing 4.65 eV light. In the final stage of generation, the 4.65 eV is combined with 1.55 eV on the third BBO to generate 6.2 eV.

The pump and probe are directed to an {optical bench} [Fig.~\ref{Fig:canopy}(b)] where the beams propagate through expanding telescopes and subsequently {are} recombined to propagate collinearly to a focusing lens. The pump and probe enter the UHV chamber through a CaF$_2$ window at 45$^\circ$ with respect {$xy$} plane of the sample [Fig.~\ref{Fig:canopy}(c)]. The expanding telescopes are used to maximize the diameter ($D$) of the beams before they are focused by a lens with focal length $f$ to minimize the 1/e$^2$ waist radius $w_0$ ($w_0 \approx 2.44 \lambda f/D$, assuming $w_0$ >> $\lambda$ and M$^2 \approx$1). In addition to minimizing $w_0$, the spacing between the lenses constituting the expanding telescope plays a vital role in defining the divergence of the beam incident on the focusing lens $f$, which, in turn, influences the focal position. We exploit this concept by incorporating a motorized translational stage capable of displacing one of the expanding telescope lenses, facilitating user control over the pump and probe 1/e$^2$ radii incident on the sample between 17 - 450 $\mu$m, and 7.9 - 200 $\mu$m, respectively.

\begin{figure}[t]
    \centering
    \includegraphics[width=1\linewidth]{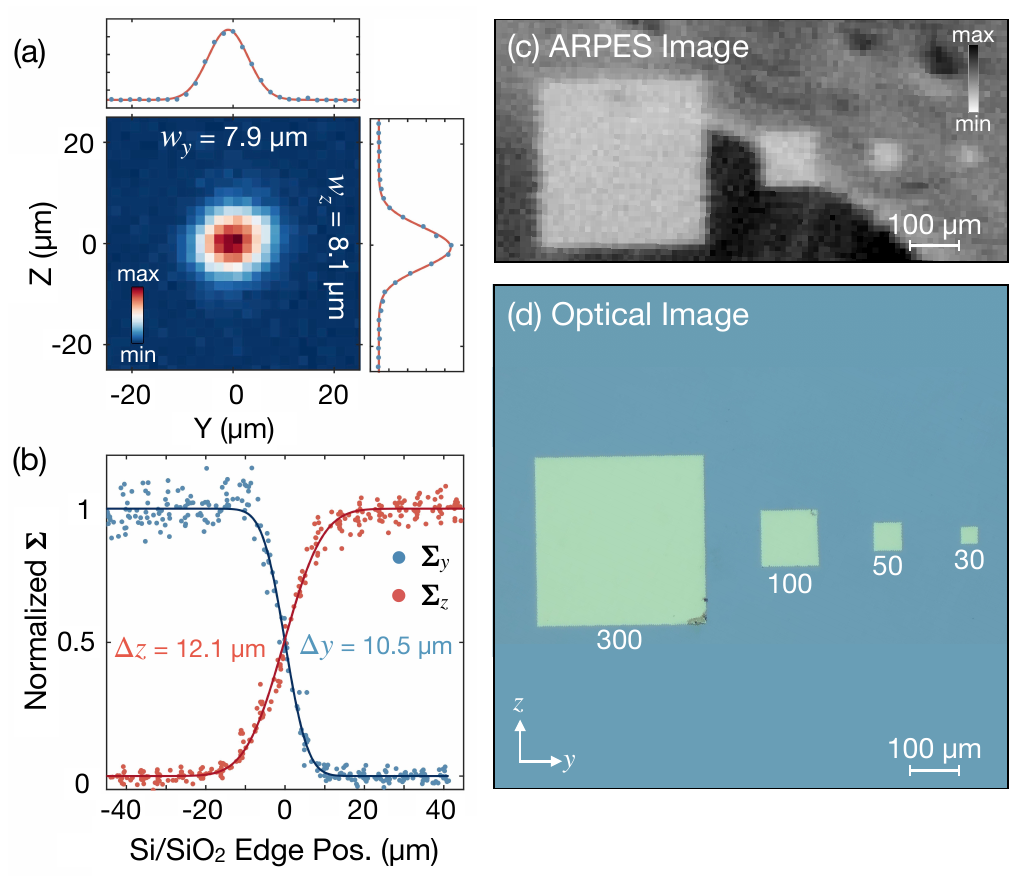}
    \caption{Spatial resolution characterization. (a) {Profile} of the 6.2 eV {probe beam} taken on a high-resolution camera equipped with a CMOS sensor showing horizontal (top panel) and vertical (right panel) cross-sectional profiles of the Gaussian beam distribution with a 1/e$^2$ waist radius of 7.9 $\mu$m and 8.1 $\mu$m, respectively. (b) Background-subtracted and normalized photoemission signal acquired across the horizontal ($\Sigma_{y}$, marked in blue) and vertical ($\Sigma_{z}$, marked in red) edge of the Si/SiO$_2$ sample, superimposed with an error function fit returning a FWHM $\Delta y$ ($\Delta z$) value of 10.5 $\pm$ 0.4 $\mu$m (12.1 $\pm$ 0.4 $\mu$m, corrected by 45$^{\circ}$ angle of incidence beam projection). (c) Two-dimensional ARPES spatial map of an Au pattern on the Si/SiO$_2$ substrate; the pixel color indicates the energy and angle-integrated photoemission signal acquired in spatial mode, binned within 10 x 10 $\mu$m$^2$ regions of the sample. (d) Corresponding optical image of the Au pattern on Si/SiO$_2$.}
    \label{Fig:squares}
\end{figure}

 {The pump and probe beam profiles are monitored by a high-resolution camera equipped with a complementary metal-oxide semiconductor (CMOS) sensor for simultaneous monitoring of the pump (800 nm) and probe (200 nm), which is positioned such that the camera sensor and sample position are equidistant from the focusing lens.} An image of the 6.2 eV beam profile {near the focal position} is shown in Fig.~\ref{Fig:squares}(a) {with a horizontal ($y$) and vertical ($z$) 1/e$^2$ radius of} {$w_{y} \approx$ 7.9 $\mu$m and $w_{z} \approx$ 8.1 $\mu$m}, respectively, as determined by a two-dimensional Gaussian fit {to the beam profile} in the form $I(y,z) \propto \exp{(-\frac{2y^2}{w_{y}^2} - \frac{2z^2}{w_{z}^2})}$. In the next section, we discuss how we combine the optical beam characterization $I(y,z)$  with the sample position {vibrations} $F(y,z)$ mentioned in Section~\ref{SystemInfo} to define the overall {effective} spatial resolution of our system, and demonstrate a method for directly measuring it using ARPES.

\begin{figure*}[t]
\centering
\includegraphics[width=.95\textwidth]{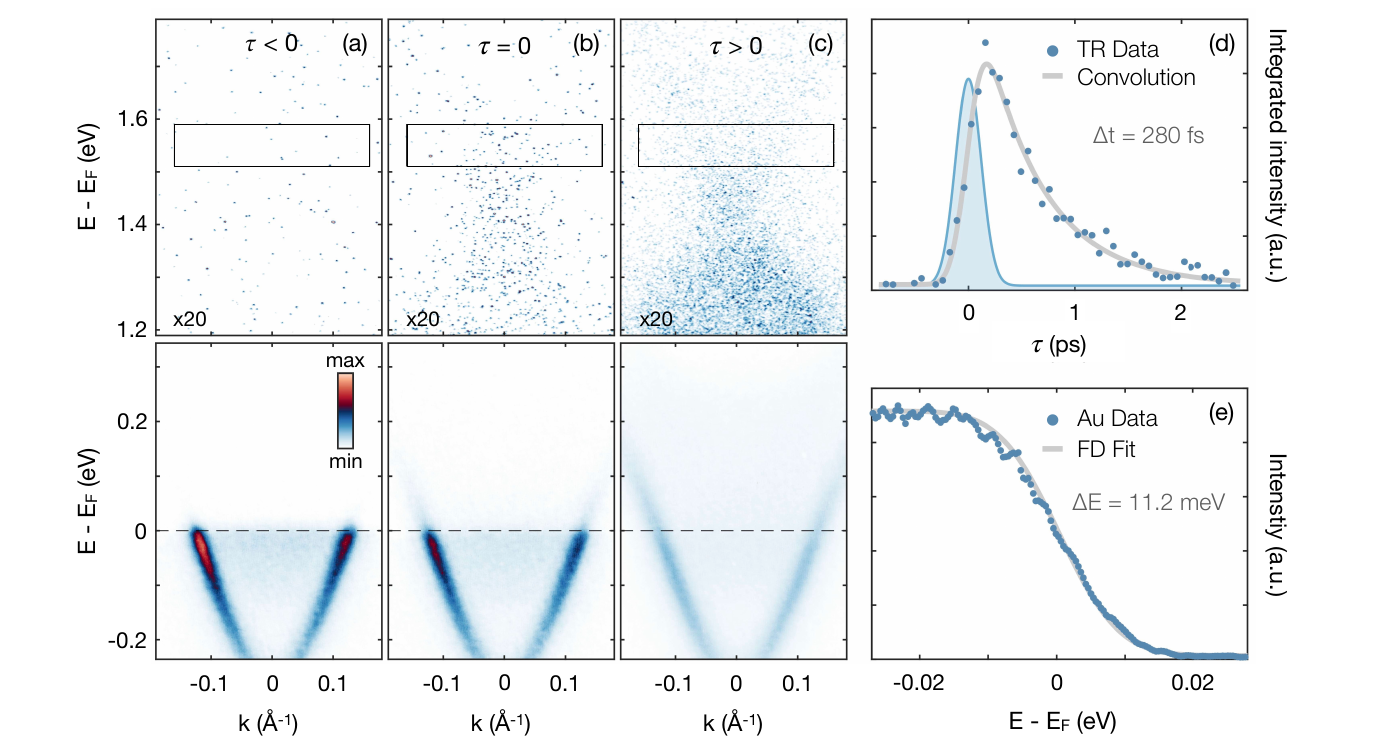}
\caption{Temporal and energy resolution characterization. (a-c) TR-ARPES intensity acquired on Bi$_2$Se$_3$ before ($\tau$ < 0), during ($\tau$ = 0), and after ($\tau$ > 0) photoexcitation by a 1.55 eV pump with an incident fluence of 0.76 mJ/cm$^2$. The top panels display the high-energy direct transition window above E$_{\text{F}}$ multiplied by a factor of 20 to be simultaneously visible with the TSS about E$_{\text{F}}$ displayed in the corresponding bottom panels. (d) The background-subtracted energy- and momentum-integrated signal (marked by blue circles), {from} the boxed region above E$_{\text{F}}$ in panels (a-c), is fit {(grey line)} to an exponential decay convolved with a Gaussian function representing the temporal  {(blue profile)}. (e) The {Fermi edge} measured {at a temperature of 10 K} on Au evaporated onto a uniform Si/SiO$_2$ substrate with the 6.2 eV light source{; t}he angle-integrated data (blue circles) are fit to a Fermi-Dirac {distribution function} convolved with a Gaussian distribution (grey), returning an energy resolution $\Delta$E = 11.2 meV.}
\label{Fig:Tres}
\end{figure*} 

\subsection{Spatial Resolution}\label{SpatialResolution}

{As mentioned in Section~\ref{SystemInfo}, the vibration of the cryostat is non-negligible on the spatial scale of the probe, and will increase the {area of the sample exposed to the probe}. To account for the contribution of sample position {vibrations} $F(y,z)$ on the spatial resolution of our system, we consider the Gaussian profile obtained {by} convolving $F(y,z)$ with the beam-profile contribution $I(y,z)$, {to obtain}: $G(y,z) \propto \exp{(-\frac{2y^2}{\sigma_{y}^2} -\frac{2z^2}{\sigma_{z}^2})}$, where {the total 1/e$^2$ radius is given by} $\sigma_{y,z} = 2 \sqrt{\delta_{y,z}^2 + (w_{y,z}/2)^2}$. Because of a combination of the beam projection on the sample, elongating the probe in the vertical direction, and a minor astigmatism of the probe, we measure the horizontal ($\Delta y$) and vertical ($\Delta z$) spatial resolutions separately, defined by the full-width at half-maximum (FWHM) of a beam with waist radii $\sigma_{y}$ ($\Delta y$ $= \sqrt{2\ln{2}} \sigma_{y}$) and  $\sigma_{z}$ ($\Delta z = \sqrt{2\ln{2}} \sigma_{z}$).}

Our approach to quantifying the spatial resolution builds on the classical "knife-edge" measurement, in which a blade eclipses a focused laser beam profile {[$G($y$) \propto \exp{(-\frac{2y^2}{\sigma_{y}^2})}$]} while a photodetector situated behind the blade measures the transmitted power, {typically in a form proportional to $[1 - \erf{(\frac{\sqrt{2}y}{\sigma_{y}})}]$,} enabling a direct measurement of the {FWHM $\Delta y$ = $\sqrt{2\ln{2}}\sigma_{y}$} ~\cite{firester_knife-edge_1977}. {To implement this measurement design in a way that is compatible with ARPES, we replace the {total transmitted power measured by a photodetector with the energy- and angle-integrated photoemission signal $\Sigma_{y,z} = \int_{\text{E}_{\text{kin}}}\int_{\textbf{k}} \text{I}_{\text{(Y,Z)}} (\text{E}_{\text{kin}},\textbf{k})\text{d}\textbf{k} \text{d}\text{E}_{\text{kin}}$ acquired in the spatial acquisition mode of the DA30-L hemispherical electron analyzer as the manipulator sweeps the horizontal ($y$) and vertical ($z$) edge of the Si/SiO$_2$ sample across the profile of the probe. Similar to the optical knife-edge method, the resulting signal $\Sigma_{y,z}$ follows an error-function distribution in both the $y$ and $z$, from which the horizontal ($\Delta y$) and vertical ($\Delta z$) spatial resolutions can be measured.}

The background-subtracted and normalized {total photoemission signal,} {$\Sigma_{y,z}$} detected across the horizontal (blue) and vertical (red) profile of the 6.2 eV beam as a function of cryostat position is shown in Fig.~\ref{Fig:squares}(b). The profiles displayed correspond to the sample-vibration-limited spatial resolution of $\Delta y$ = 10.5 $\pm$ 0.4 $\mu$m, and {$\Delta z$ = 12.1 $\pm$ 0.4 $\mu$m} (projection corrected). In an effort to provide a straightforward indicator of the overall spatial resolution of our system, we define the effective spatial resolution $\Delta_{\textit{eff}}=\sqrt{\Delta y \Delta z}$ as the FWHM of a 2D Gaussian with an area equal to that enclosed by an ellipse with radii $\Delta y/2$ and $\Delta z/2$, resulting in $\Delta_{\textit{eff}}$ = 11.3 $\pm$ 0.3 $\mu$m.

For a second test, we fabricate a structure that can be measured both optically and by photoemission for a direct comparison. Here, the sample consists of a Si/SiO$_2$ wafer, covered with a 5 nm layer of evaporated Ti to enhance the adherence of the 45 nm thick Au layer in a cascading-squares {pattern} (from 300 - 30 $\mu$m) fabricated by photo-lithography, as illustrated in the Fig.~\ref{Fig:canopy}(c). A high-resolution optical image of the sample in UHV is shown in Fig.~\ref{Fig:squares}(d){,} for comparison with the ARPES image [Fig.~\ref{Fig:squares}(c)] obtained by raster scanning the sample across the profile of the 6.2 eV {beam}, and plotting the $\Sigma_{y,z}$ value obtained in spatial mode binned in 10 $\times$ 10 $\mu$m$^2$ regions. This map serves to highlight the imaging capability {of} our system, as we are able to locate and acquire meaningful spectral information from samples smaller than 30 $\mu$m.

\subsection{Temporal Resolution}\label{TemporalRes}

To assess the temporal resolution of our instrument ($\Delta$t) and fine-tune the spatial overlap of the pump and probe, we use the topological insulator Bi$_2$Se$_3$, of which the photoinduced dynamics has been thoroughly characterized~\cite{sobota_ultrafast_2012,sobota_ultrafast_2014}. The temporal resolution is defined as the {duration of the} cross-correlation {between} the probe and pump pulses given by $\Delta \text{t} = \sqrt{(\Delta \text{t}_{\text{probe}})^2 + (\Delta \text{t}_{\text{pump}})^2}$, where $\Delta \text{t}_{\text{probe}}$ ($\Delta \text{t}_{\text{pump}}$) is the pulse duration of the probe (pump), and is measured experimentally by {tracking the relaxation dynamics} of the direct-transition population at $\sim$1.55 eV above E$_{\text{F}}$. For this measurement, Bi$_2$Se$_3$ was cleaved {\textit{in-situ}} at 10 K and pumped with p-polarized 1.55 eV {photons} at an incident fluence of 0.76 mJ/cm$^2$; the equilibrium spectra of the topological surface state (TSS) are shown in the bottom panel of Fig.~\ref{Fig:Tres}(a), where the Dirac-point is located $\sim$300 meV below E$_{\text{F}}$. 

In {the top panels of} Fig.~\ref{Fig:Tres}(a-c), we show the evolution of the direct transition population into the electronic states more than 1.55 eV above E$_{\text{F}}$. The {time}-resolution-limited onset and subsequent relaxation of the direct transient population {1.55 eV above E$_{\text{F}}$ [boxed region in the top panels of Fig.~\ref{Fig:Tres}(a-c)] are {shown} by blue circles in Fig.~\ref{Fig:Tres}(d), and fit to an exponential decay convolved with a Gaussian function (solid grey line); the FWHM of the Gaussian function [shaded blue region in Fig.~\ref{Fig:Tres}(d)] returns a {total} temporal resolution of $\sim$280 fs}. {This cross-correlation of pump and probe pulses, together with the earlier {autocorrelation} determination of $\text{t}_{\text{pump}} \sim$150 fs [Fig.~\ref{Fig:canopy}(d)], assuming a Gaussian pulse shape, allows us to extract $\text{t}_{\text{probe}} \sim$ 236 fs.}

\subsection{Energy Resolution}\label{EnergyRes}

The total energy resolution of the system is a combination of the analyzer contribution $\Delta \text{E}_{\text{ana}}$ and the bandwidth of the laser $\Delta \text{E}_{\text{h}\nu}$. To measure the total energy resolution, we use a Si/SiO$_2$ wafer fabricated using the method described in Section~\ref{SpatialResolution}, and uniformly cover the surface \textit{in-situ} by evaporating Au at 10 K. ARPES spectra were collected using the 6.2 eV {photons} with a straight analyzer entrance slit width of 0.05 mm and pass energy of 5 eV. The resulting angle-integrated spectrum is shown in Fig.~\ref{Fig:Tres}(e). We fit this measured intensity to a Fermi-Dirac distribution, with a width proportional to k$_\text{B}$T, convolved with a Gaussian function, representing the total energy resolution, $\Delta \text{E}$, which was measured to be $\sim$11.2 meV. Assuming $\Delta \text{E}_{\text{h}\nu}$ >> $\Delta \text{E}_{\text{ana}}$, {from the previously determined $\text{t}_{\text{probe}} \sim$236 fs} the upper limit of the time-energy bandwidth product of our system is {2643} meV$\cdot$fs, which is $\sim$1.4 times the transform-limited time-bandwidth product for a Gaussian-shaped pulse{,which has a minimum value of $\Delta$E$\cdot \Delta$t$\geq 4\hbar \ln{2} \approx 1825$ mev$\cdot$fs}~\cite{gauthier_tuning_2020}.

\section{B\lowercase{i}$_2$S\lowercase{e}$_3$ Spatial Resolution}\label{Bi2Se3}

 \begin{figure}[t]
    \centering
    \includegraphics[width=0.9\linewidth]{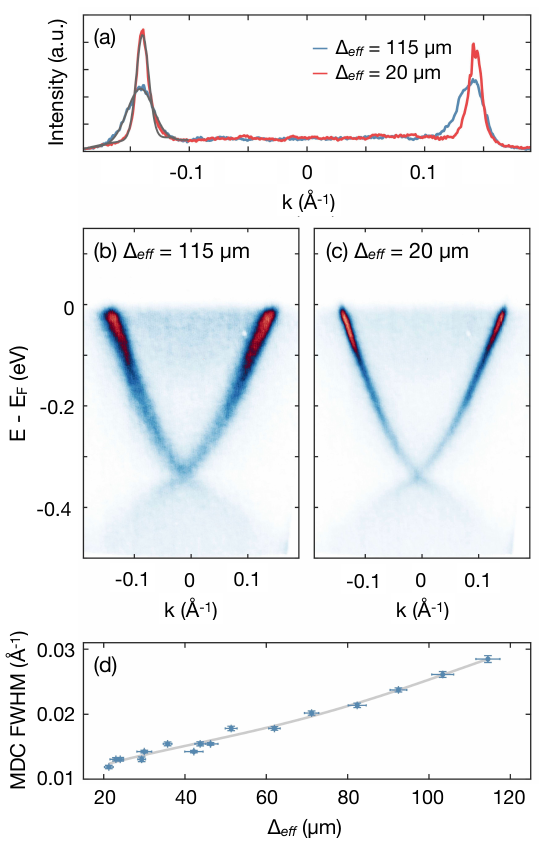}
    \caption{{Spot-size dependent spectral featured of Bi$_2$Se$_3$}. (a) MDCs extracted from a 15 meV wide region near E$_{\text{F}}$ in two regimes: large spatial resolution $\Delta_{\textit{eff}} \approx$ 115 $\mu$m (blue solid line), and small $\Delta_{\textit{eff}} \approx$ 20 $\mu$m (red solid line). {Fits including a } Gaussian plus polynomial background are shown in grey. (b,c) ARPES spectra of Bi$_2$Se$_3$ acquired with a $\Delta$$_{\textit{eff}}$ of 115 $\mu$m and 20 $\mu$m, respectively. (d) MDC linewidth as a function of the total SR area (area enclosed in an illuminated region with a diameter of $\Delta_{\textit{eff}}$).} 
    \label{Fig:bise}
\end{figure}

The spectral features {measured by} ARPES inherently possess a finite linewidth arising from fundamental many-body interactions, where increased electronic scattering reduces lifetime {and}, consequently{, leads to} broadening~\cite{damascelli_angle-resolved_2003,iwasawa_high-resolution_2020}. However, before drawing conclusions about electronic interactions from these spectral features, {it is} essential to consider how extrinsic factors contribute to spectral broadening, and may potentially overshadow the {intrinsic} linewidth. These extrinsic factors include angular uncertainty introduced by sample surface inhomogeneity (originating from inhomogeneous chemical doping, or the presence of several domains, for example), and stray field effects arising from a large difference in the work function between the sample and the substrate {or surroundings}, which are likely to distort the {path of} low-kinetic energy electrons~\cite{fero_impact_2014}. While the primary motivation for developing a $\mu$-ARPES setup is to investigate the electronic properties of exfoliated crystalline materials, another objective is to enhance data quality obtained from large samples by minimizing the contribution of extrinsic factors that overwhelm inherent spectral linewidth measured. {Here, by taking advantage of the full control of the effective spatial resolution, we show that microscopic inhomogeneities of the macroscopically flat surface of the well-studied Bi$_2$Se$_3$ topological insulator lead to a considerable broadening of the ARPES lineshape.}

Measurements on Bi$_2$Se$_3$ {cleaved \textit{in-situ}} were performed at 10 K while systematically varying the size of the 6.2 eV probe to obtain ARPES spectra corresponding to total spatial resolution $\Delta_{\textit{eff}}$ values between $\sim$20 - 115 $\mu$m. Spectra acquired with $\Delta_{\textit{eff}}$ values of $\sim$115 and 20 $\mu$m are depicted in Fig.~\ref{Fig:bise}(b) and (c), respectively. For a direct comparison between these spectra, the momentum distribution curves (MDCs) at E$_{\text{F}}$ are shown in Fig.~\ref{Fig:bise}(a){, superimposed with a Gaussian fit plus polynomial background (solid grey line), from which the FWHM of the TSS near E$_{\text{F}}$ is extracted}. The {spectral width of the TSS, as determined by the FWHM from the composite fit to the MDCs,} increases from $\sim$0.012 $\angstrom^{-1}$ to $\sim$0.029 $\angstrom^{-1}$, more than double, when the $\Delta_{\textit{eff}}$ is increased from 20  $\mu$m to 115  $\mu$m, demonstrating how strongly the measured spectral linewidth can be tied to extrinsic sources, rather than intrinsic factors. {To further demonstrate this dependence, we extract the width of the TSS for $\Delta$$_{\textit{eff}}$ values ranging from 20 to 115$\mu$m by systematically changing the spot size of the probe incident on the sample; the FWHM of the MDCs as a function of spatial resolution {is} shown in Fig.~\ref{Fig:bise}(d), where we report a clear correlation between the spectral broadening and probe spot-size.}

\begin{figure}[t]
    \centering
    \includegraphics[width=1.0\linewidth]{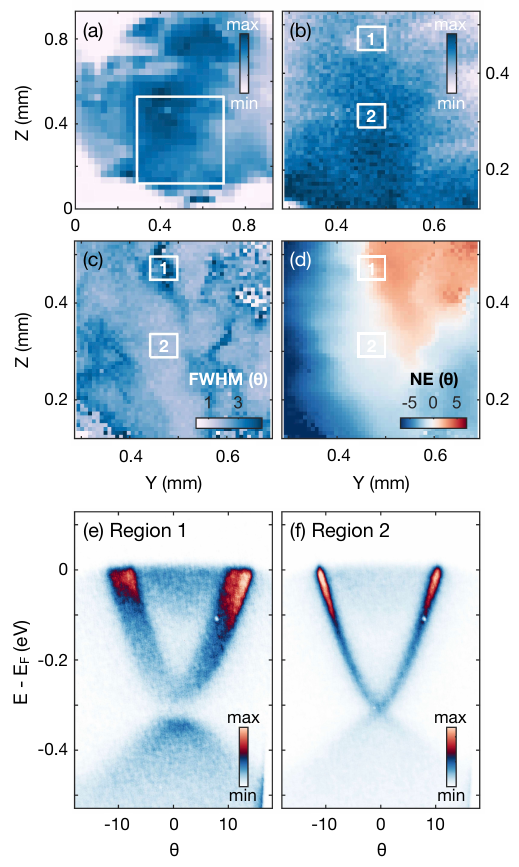}
    \caption{Spatially-resolved electronic structure of Bi$_2$Se$_3$. (a) Photoemission intensity of the TSS within 70 meV of E$_{\text{F}}$ as a function of the spatial position Y and Z. (b) High resolution (5 $\mu$m step size) grid scan of the region outlined in the white box in panel (a). (c) and (d) {average MDC FWHM at E$_{\text{F}}$ and {relative} angle of normal emission (NE) as a function of the sample position}. (e,f) ARPES spectra collected within the sample region outlined in boxes 1 and 2, respectively, in panels (b-d).} 
    \label{Fig:bisegrid}
\end{figure}

Before investigating the {precise} origin of this {extrinsic} spectral broadening {becoming more prominent the larger the beam spot}, we will briefly mention one caveat to the $\mu$-ARPES approach, {namely} the ever-increasing contribution of space-charge effects which can {in principle} be an {additional} source of {extrinsic} sample broadening. A tightly-focused pulsed light source can emit a dense cloud of spatially and temporally confined photoelectrons that experience significant Coulomb repulsion, leading to {spatial spreading and kinetic energy broadening}~\cite{hellmann_vacuum_2009,graf_vacuum_2010}. {This is however not a concern here:} in the context of the results shown in Fig.~\ref{Fig:bise}, if space-charge effects were contributing we should have reported a {progressive} increase in the linewidth as we increase the probe fluence upon {reducing the beam spot, at variance with what we observed experimentally}. Additionally, space charge effects {were} mitigated prior to measurements by reducing the probe spot size to the minimum, and subsequently attenuating the probe until no shift of the Fermi-level is observed {with the tightest possible focus}.

\begin{figure*}[t!]
\centering
\includegraphics[width=\textwidth]{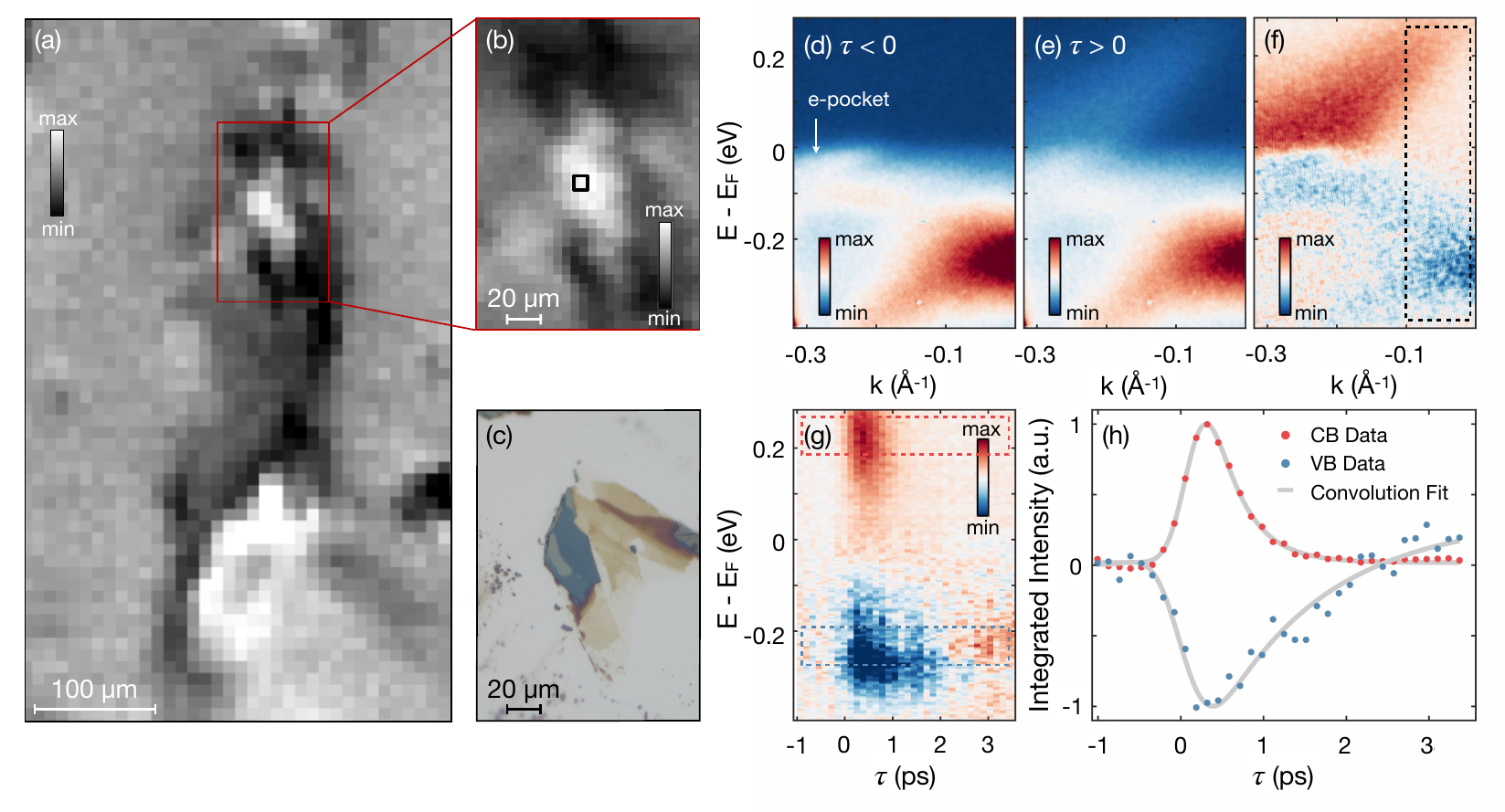}
\caption{{Time-resolved $\mu$-ARPES on WTe$_2$.} (a) Large two-dimensional {map} of WTe$_2$ flakes exfoliated onto Ag(111) {acquired by $\mu$-ARPES in 30 $\mu$m steps}, {with the color map representing the momentum and energy integrated photoemission signal within a 200 meV region below E$_{\text{F}}$.} (b) High-resolution two-dimensional $\mu$-ARPES {map of the region boxed in red in panel (a) obtained with a} 5 $\mu$m step size, and (c) corresponding optical image of WTe$_2$ on Ag(111){, where the strength of the optical contrast to the grey substrate is indicative of sample-thickness}. {(d,e) TR-ARPES spectra acquired before ($\tau$ < 0) and after ($\tau$ > 0) photoexcitation with a 1.55 eV pump with an incident fluence of 0.62 mJ/cm$^2$, respectively; (f)  differential TR-ARPES map of the spectra shown (d) and (e). (g) Transient momentum-integrated [k$_{||}$ > -0.1 $\angstrom^{-1}$, indicated by the black dashed box in panel (f)] energy distribution curves from the {difference} spectra in the black dashed box in panel (f). (h) Time traces obtained from panel (g) highlighting the transient population of the conduction band [marked by the red dashed box in panel (g)], and suppression and subsequent recovery of the valance band population [blue box in (g)].}}
\label{Fig:WTe2}
\end{figure*} 

To {further} investigate {extrinsic} contributions to the spectral broadening, we use the tightly-focused 6.2 eV photons in combination with high-resolution sample scanning to resolve surface inhomogeneities {that could be potentially contributing to the observations {presented} in Fig.~\ref{Fig:bise}}. In Fig.~\ref{Fig:bisegrid}(a),(b) we show an ARPES spatial map of the entire cleaved Bi$_2$Se$_3$ sample [Fig.~\ref{Fig:bisegrid}(a)], and a high-resolution scan of the subsection [Fig.~\ref{Fig:bisegrid}(b)] enclosed in the white boxed region in Fig.~\ref{Fig:bisegrid}(a). {Within} each {30 $\times$ 30 $\mu$m$^2$ (5 $\times$ 5 $\mu$m$^2$)} pixel in Fig.~\ref{Fig:bisegrid}(a) [Fig.~\ref{Fig:bisegrid}(b)], {ARPES spectra are acquired in angular mode, and the photoemission signal within 70 meV of the Fermi level is integrated over all angles ($\pm$ 20$^{\circ}$), which defines the color scale of this image.} The high-resolution {intensity} map in Fig.~\ref{Fig:bisegrid}(b) appears {macroscopically} uniform, with little variation in the {integrated} intensity observed over this region of the sample. However, variations become apparent when the FWHM of the TSS close to E$_{\text{F}}$ [Fig.~\ref{Fig:bisegrid}(c)] and {relative} {angle of normal emission} [Fig.~\ref{Fig:bisegrid}(d)] is extracted for each 5 x 5 $\mu$m$^2$ region shown in Fig.~\ref{Fig:bisegrid}(b).

{Even with an inhomogeneous sample surface such as the one displayed here, we are able to obtain high quality and sharp spectral features [Fig.~\ref{Fig:bisegrid}(e), for example] by reducing the probe size to {dimensions} smaller than the surface variations.} Additionally, by minimizing the contribution of extrinsic sources, we begin to approach a regime where the spectral broadening is {representative of} the intrinsic many-body interactions, {as} shown by the gradual trend towards sharper linewidths observed in Fig.~\ref{Fig:bise}(d).

\section{WT\lowercase{e}$_2$}\label{WTe2}

Beyond {resolving} surface inhomogeneity, the small-spot capability of our system is useful for studying ultrafast dynamics in exfoliated materials and stacked/twisted layers on the order of tens of $\mu$m. In this section, we use \text{in-situ} exfoliated transition metal dichalcogenide (TMD) WTe$_2$ to showcase our time-resolved $\mu$-ARPES apparatus as a method for measuring the electronic properties of exfoliated materials. WTe$_2$ is a type-II Weyl semimetal exhibiting large magnetoresistance~\cite{tang_quantum_2017,shi_imaging_nodate,das_electronic_2019}, and has previously been the subject of extensive TR-ARPES and $\mu$-ARPES studies owing to its potential as a platform for the study of topological quantum effects~\cite{cucchi_microfocus_2019,hein_mode-resolved_2020,caputo_dynamics_2018,wu_observation_2016}. {In addition, successful exfoliation in UHV of WTe$_2$ via} kinetic \textit{in-situ} single-layer synthesis (KISS) technique has been reported~\cite{grubisic_cabo_situ_2023}, making WTe$_2$ a good candidate for a demonstration of the performance of our time-resolved $\mu$-ARPES setup for the study of ultrafast dynamics in exfoliated materials. 

A {400 x 400 $\mu$m$^2$} Ag(111) substrate was prepared by 1 keV Ar-ion sputtering and 550 K annealing. Bulk WTe$_2$ was cleaved \textit{in-situ}, revealing an atomically clean surface, and exfoliated onto the Ag(111) substrate using the KISS method{, which consists of stamping the clean Ag(111) substrate with the surface of the cleaved bulk crystal}. Flakes of WTe$_2$ were located by scanning the sample in 30 $\mu$m steps as shown in Fig.~\ref{Fig:WTe2}(a), where the color scale represents the {energy and momentum integrated} photoemission intensity {within} a region $\sim$200 meV below E$_{\text{F}}$.  A {high-resolution (5 $\mu$m)} {spatial} scan of this flake is shown in Fig.~\ref{Fig:WTe2}(b) with a direct comparison to the corresponding optical contrast image in Fig.~\ref{Fig:WTe2}(c), {revealing a $\sim$30 $\mu$m flake of WTe$_2$}. {The sample} was orientated along the {$\Gamma-X$} direction and probed with p-polarized 6.2 eV photons, as shown in Fig.~\ref{Fig:WTe2}(d), where we observe the electron pocket {centered} around k$_{||}$ = -0.3 $\angstrom^{-1}$. Figure~\ref{Fig:WTe2}(f) displays the differential ARPES map [I($\tau$ > 0) - I($\tau$ < 0)], where we observe a transient enhancement in the population of the electron pocket above E$_{\text{F}}$ (red area) along with a suppression of spectral weight in the valence band below E$_{\text{F}}$ (blue area), consistent with previous observations ~\cite{hein_mode-resolved_2020,bruno_observation_2016,caputo_dynamics_2018}. 

To further highlight the transient dynamics, the momentum-integrated energy distribution curves (EDCs) are exacted from the differential map, corresponding to the region boxed in Fig.~\ref{Fig:WTe2}(f) (k$_{||}$ > -0.1 $\angstrom^{-1}$), and displayed as a function of pump-probe delay in Fig.~\ref{Fig:WTe2}(g). Here, the transient enhancement of the conduction band, and depletion in the valance band are extracted from energy regions centered at E-E$_{\text{F}}$ = $\pm$0.23 eV as depicted by the dashed boxed in Fig.~\ref{Fig:WTe2}(g), and shown in Fig.~\ref{Fig:WTe2}(h). {The transient dynamics of the conduction and valence bands are well captured by an exponential decay (convolved with a Gaussian{ of $\Delta$t $\approx 300$ fs}) of 0.38 ps and 1.34 ps, respectively, in good agreement with previous reports~\cite{hein_mode-resolved_2020}.} 

{These results demonstrate that} the combination of microscale and time-resolved ARPES opens the gate to a myriad of new experimental applications, {ushering} the study of two-dimensional materials into the time domain.

\section{Conclusion}

In this work, we have presented the capabilities of our time-resolved $\mu$-ARPES system for characterizing the electronic properties of exfoliated materials in both equilibrium and photo-excited conditions. With its combination of tunable vibration-limited spatial resolution, tunable pump fluence, high energy resolution, and near-transform-limited temporal resolution, our system is well-suited for probing intrinsic electron interactions and dynamics from spectral features on $\mu$m-sized quantum materials. {Indeed, we have shown how spatial inhomogeneities on the $\mu$m-scale impact the ARPES lineshape of the topological surface state of Bi$_2$Se$_3$.} In addition, we have demonstrated the system's ability to locate and measure exfoliated TMDs after \textit{in-situ} exfoliation. These capabilities make our time-resolved $\mu$-ARPES system a valuable tool for investigating the ultrafast dynamics and advancing our understanding of their emerging electronic properties.

\begin{acknowledgments}
We gratefully acknowledge O. Toader for help with cryostat stabilization, {P. Sulzer and M. X. Na for insightful discussions}, and M. Kroug, A. Blednov, and K. Michelakis for Si/SiO$_2$/Au photolithography sample fabrication. {This project was supported through the CIFAR Catalyst award "Electronic structure and dynamics of microscale quantum materials \& devices".} This research was undertaken thanks in part to funding from the Max Planck-UBC-UTokyo Centre for Quantum Materials and the Canada First Research Excellence Fund, Quantum Materials and Future Technologies Program. This project is also funded by the Gordon and Betty Moore Foundations EPiQS Initiative, Grant No. GBMF4779 to A.D. and D.J.J.; the Natural Sciences and Engineering Research Council of Canada (NSERC); the Canada Foundation for Innovation (CFI); the British Columbia Knowledge Development Fund (BCKDF); the Department of National Defence (DND); the Canada Research Chairs Program (A.D.); and the CIFAR Quantum Materials Program (A.D.).
\end{acknowledgments}



\clearpage
%

\end{document}